\documentclass[seceq]{ptptex}

\usepackage{graphicx}

\title{Strange meson production at high density and temperature
}

\author{
Laura \textsc{Tolos}$^1$, Daniel \textsc{Cabrera}$^2$, Raquel \textsc{Molina}$^3$, Eulogio \textsc{Oset}$^3$, \\ Angels \textsc{Ramos}$^4$}

\inst{$^1$Theory Group. KVI. University of Groningen,
Zernikelaan 25, 9747 AA Groningen, The Netherlands \\
$^2$Departamento de F\'{\i}sica Te\'orica II, Universidad Complutense,\\
28040 Madrid, Spain\\
$^3$Instituto de F{\'\i}sica Corpuscular (centro mixto CSIC-UV),
Institutos de Investigaci\'on de Paterna, Aptdo. 22085, 46071, Valencia, Spain \\
$^4$Departament d'Estructura i Constituents de la Mat\`eria,
Universitat de Barcelona,
Diagonal 647, 08028 Barcelona, Spain}

\abst{
The properties of strange mesons ($K$, $\bar K$ and $\bar K^*$) in dense matter are studied using a unitary approach in coupled channels for meson-baryon scattering. The kaon-nucleon interaction incorporates $s$- and $p$-wave contributions within a chiral model whereas the interaction of $\bar K^*$ with nucleons is obtained in the framework of the local hidden gauge formalism. The in-medium solution for the scattering amplitude accounts for Pauli blocking effects, mean-field binding on baryons, and meson self-energies. We obtain the $K$, $\bar K$ and $\bar K^*$ (off-shell) spectral functions in the nuclear medium  and study their behaviour at finite density, temperature and momentum. We also analyze the energy weighted sum rules of the kaon propagator as a quality test of model calculations. We finally estimate the transparency ratio of the $\gamma A \to K^+ K^{*-} A^\prime$ reaction, which we propose as a feasible scenario at present
facilities to detect in-medium modifications of the $\bar K^*$ meson.
}

\begin{document}

\maketitle
\section{Introduction}

Strangeness in hot and dense matter is a matter of extensive analysis in connection to heavy-ion collisions from SIS \cite{Fuchs:2005zg} to FAIR \cite{fair} energies at GSI. In particular, the interaction of strange pseudoscalar mesons ($K$ and $\bar K$) with matter is a topic of high interest. Whereas the interaction of $\bar K N$ is repulsive at threshold, the phenomenology of antikaonic atoms \cite{Friedman:2007zz} shows that the $\bar K$ feels an attractive potential at low densities. This attraction is a consequence of the modified $s$-wave $\Lambda(1405)$ resonance in the medium due to Pauli blocking effects \cite{Koch} together with the self-consistent consideration of the $\bar K$ self-energy \cite{Lutz} and the inclusion of self-energies of the mesons and baryons in the intermediate states \cite{Ramos:1999ku}. Attraction of the order of -50 MeV at normal nuclear matter density, $\rho_0=0.17 \,{\rm fm^{-3}}$, is obtained in unitarizated theories in coupled channels based on chiral dynamics \cite{Ramos:1999ku} and meson-exchange models \cite{Tolos01,Tolos02}. Moreover, the knowledge of higher-partial waves beyond $s$-wave \cite{Tolos:2006ny,Lutz:2007bh,Tolos:2008di} becomes essential for relativistic heavy-ion experiments at beam energies below 2AGeV \cite{Fuchs:2005zg}.

Curiously, no discussion has been made about the properties of the strange
vector mesons ($K^*$ and $\bar K^*$) in the medium, although non-strange vector mesons have been the focus of attention for years \cite{rapp,hayano,mosel}. Only recently the $\bar K^*N$ interaction in free space has been addressed
in  \cite{GarciaRecio:2005hy} using SU(6) spin-flavour symmetry, and in \cite{Oset:2009vf} within the hidden local gauge formalism for the interaction of vector mesons with baryons of the octet and the decuplet. Within this latter scheme, medium effects have been implemented and analyzed very recently \cite{tolos10} finding  
an spectacular enhancement of the $\bar{K}^*$ width in the medium, up to about five times the free value of about $50~{\rm MeV}$.

In this paper we review the properties of the strange ($K$, $\bar K$ and $\bar K^*$) mesons in dense matter. We then analyze the first energy-weighted sum rules (EWSRs) for kaons, which are a quality test of our model calculations. We finally estimate the transparency ratio of the $\gamma A \to K^+ K^{*-} A^\prime$ reaction, as a feasible scenario at present facilities to detect in-medium properties of strange vector mesons.

\section{$K$ and $\bar K$ mesons in hot and dense matter}

The  kaon and antikaon self-energies in symmetric nuclear matter at finite
temperature are obtained from the $s$- and $p$-waves in-medium kaon-nucleon interaction within a chiral unitary approach \cite{Tolos:2008di}.

 The $s$-wave amplitude of the $\bar K N$ originates, at tree level, from the Weinberg-Tomozawa term of the chiral Lagrangian. Unitarization in coupled channels is imposed with on-shell amplitudes ($T$) and a cutoff regularization. The $\Lambda(1405)$ resonance in the $I=0$ channel is generated dynamically and we obtain a satisfactory description of low-energy scattering observables. The $K N$ effective interaction is also obtained by solving the Bethe-Salpeter equation with the same cutoff parameter. 

 The in-medium solution of the $s$-wave amplitude accounts for Pauli-blocking effects, mean-field binding on the nucleons and hyperons via a $\sigma-\omega$ model, and the dressing of the pion and kaon propagators. The self-energy is then obtained in a self-consistent manner summing the transition amplitude $T$ for the different isospins over the nucleon Fermi distribution at a given temperature, $n(\vec{q},T)$,  
\begin{eqnarray}
\Pi_{K(\bar K) N}(q_0,{\vec q},T)= \int \frac{d^3p}{(2\pi)^3}\, n(\vec{p},T) \,
[\, {T}_{K(\bar K)N}^{(I=0)} (P_0,\vec{P},T) +
3 \, {T}_{K(\bar K)N}^{(I=1)} (P_0,\vec{P},T)\, ], \ \  \ \ \ \label{eq:selfd}
\end{eqnarray}
where $P_0=q_0+E_N(\vec{p},T)$ and $\vec{P}=\vec{q}+\vec{p}$ are
the total energy and momentum of the kaon-nucleon pair in the nuclear
matter rest frame, and ($q_0$, $\vec{q}\,$) and ($E_N$, $\vec{p}$\,) stand  for
the energy and momentum of the kaon and nucleon, respectively, also in this
frame. In the case of $\bar K$ meson the model also includes, in addition, a $p$-wave contribution to the self-energy from hyperon-hole ($Yh$) excitations, where $Y$ stands for $\Lambda$, $\Sigma$ and
$\Sigma^*$ components. For the $K$ meson the $p$-wave self-energy results from $YN^{-1}$ excitations in crossed kinematics. The spectral function depicted in the following results from the imaginary part of the in-medium kaon propagator.

\begin{figure}[htb]
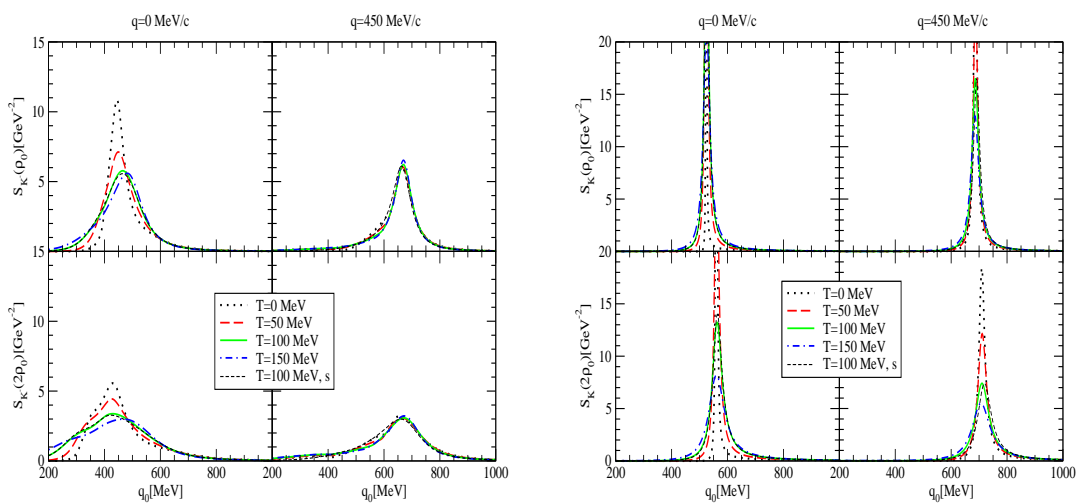

\begin{center}
\includegraphics[height=6.5 cm, width=6.6 cm]{fig6.eps}
\hfill
\includegraphics[height=6.5 cm, width=6.6 cm]{fig7.eps}
\caption{ $\bar K$ (left) and $K$ (right) spectral functions for different densities, temperatures and momenta.}
 \label{fig1}
\end{center}
\end{figure}

The evolution of the  $\bar{K}$ and $K$ spectral functions with density and temperature is depicted in Fig.~\ref{fig1}. The $\bar K$ spectral function (left) shows a strong mixing between the quasi-particle peak and the $\Lambda(1405)N^{-1}$ and  $Y(=\Lambda, \Sigma , \Sigma^*)N^{-1}$ excitations. The effect of these $p$-wave $YN^{-1}$ subthreshold excitations is
repulsive for the $\bar K$ potential, compensating in part the attraction
from the $s$-wave ${\bar K} N$ interaction.  Temperature  and density softens the
$p$-wave contributions to the spectral function at the quasi-particle energy. Moreover, together with the $s$-wave, the $p$-wave self-energy
provides a low-energy tail which spreads the spectral function considerably. As for the $K$ spectral function (right), the $K$ meson is described by a narrow quasi-particle peak which dilutes 
with temperature and density as the phase space for 
$KN$ states increases. The $s$-wave repulsive
self-energy  translates into a shift of the $K$ spectral function to
higher energies with increasing density and it is mildly compensated by the attractive $p$-wave contributions.

\subsection{Energy weighted sum rules for kaons}

The EWSRs result from matching the Dyson form of the meson propagator with its spectral Lehmann representation at low and high 
energies \cite{ewsr}. The first EWSRs in the high-energy limit, $m_0^{(\mp)}$, together with the zero energy EWSR, $m_{-1}$, are 
\begin{eqnarray}
m_{-1}&:&
\int_0^{\infty} \textrm{d}\omega \,
\frac{1}{\omega} \, [ S_{\bar K}(\omega,\vec{q}\,;\rho,T) + S_{K}(\omega,\vec{q}\,;\rho,T)]
=
\frac{1}{\omega_{\bar K}^2(\vec{q}\,) + \Pi_{\bar K}(0,\vec{q}\,;\rho,T)} \ \ \ \  
\\
m_{0}^{(\mp)}&:&  
\int_0^{\infty} \textrm{d}\omega \,
[ S_{\bar K}(\omega,\vec{q}\ ;\rho,T) - S_{K}(\omega,\vec{q}\,;\rho,T) ] = 0
\nonumber \\
& & 
\int_0^{\infty} \textrm{d}\omega \,
\omega \, [ S_{\bar K}(\omega,\vec{q}\,;\rho,T) + S_{K}(\omega,\vec{q}\,;\rho, T) ] = 1 \ .
\end{eqnarray}

\begin{figure}[htb]
\begin{center}
\includegraphics[height=8 cm, width=6 cm]{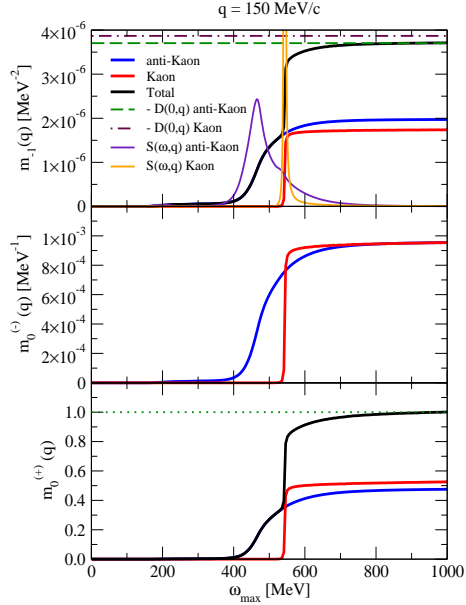}
\caption{ $m_{-1}$, $m_{0}^{(-)}$ and $m_{0}^{(+)}$ sum rules for the $K$ and $\bar K$ spectral functions at
$q=150$~MeV/c, $\rho=\rho_0$ and $T=0$ MeV. The $\bar K$ and $K$ spectral
functions are also displayed for reference.}
 \label{fig2}
\end{center}
\end{figure}

In Fig.~\ref{fig2} we show the sum rules for the antikaon propagator as a function of the upper integral limit for
$\rho=\rho_0$, $T=0 \ {\rm MeV}$ and $q=150~$MeV/c. The contributions from $\bar K$ and $K$ to the l.h.s. of the sum rule are depicted separately. The $\bar K$ and $K$ spectral functions are also shown for reference in arbitrary units. Note that saturation is progressively shifted to higher energies as we examine sum rules involving higher order weights in energy.

The l.h.s. of the $m_{-1}$ sum rule (upper panel) saturates satisfactorily a few hundred MeV beyond the quasiparticle peak, following the behaviour of the $\bar K$ and $K$ spectral functions. The $m_0^{(-)}$ sum rule shows that the areas subtended by the $K$ and $\bar K$ spectral functions coincide (middle panel). The fullfilment of this sum rule is, however, far from trivial because, although the $\bar K$ and $K$ spectral
functions are related by the retardation property,  the actual calculation of the meson self-energies is
done exclusively for positive meson energies. Finally, the $m_0^{(+)}$ sum rule (lower panel) saturates to one independently of the meson momentum, nuclear density or temperature, thus posing a strong constraint on the accuracy of the calculations. Moreover, those sum rules have been also tested satisfactorily for higher momenta and temperature \cite{ewsr}.

\section{$\bar K^*$ meson in nuclear medium}

The  $\bar K^*$  self-energy in symmetric nuclear matter is obtained  within the hidden gauge formalism \cite{tolos10}. There are two sources for the modification of the $\bar K^*$ $s$-wave self-energy in nuclear matter: a) the contribution associated to the decay mode $\bar K \pi$ modified by the nuclear medium effects on the $\bar K$ and $\pi$ mesons, and b) the contribution associated to the interaction of the $\bar K^*$ with the nucleons in the medium, which accounts for the direct quasi-elastic process $\bar K^* N \to \bar K^* N$ as well as other absorption channels $\bar K^* N\to \rho Y, \omega Y, \phi Y, \dots$ with $Y=\Lambda,\Sigma$. In fact, this last term comes from a unitarized coupled-channel process, similar to the $\bar K N$ case. Two resonances are generated dynamically, 
 $\Lambda(1783)$ and  $\Sigma(1830)$, which can be identified with the experimentally observed states $J^P=1/2^-$ $\Lambda(1800)$ and the $J^P=1/2^-$ PDG state $\Sigma(1750)$, respectively \cite{Oset:2009vf}.

The in-medium $\bar K^*$ self-energy results from the sum of both contributions,
$\Pi_{\bar K^*}=\Pi_{\bar{K}^*}^{{\rm (a)}}
+\Pi_{\bar{K}^*}^{{\rm (b)}}$,
where $\Pi_{\bar{K}^*}^{{\rm (b)}}$ is obtained similarly to $\bar K N$  by integrating the $\bar K^* N$ transition amplitude over the nucleon Fermi sea,
\begin{eqnarray}
\Pi_{\bar{K}^*}^{{\rm (b)}}(q_0,\vec{q}\,)&=&\int \frac{d^3p}{(2\pi)^3} \, n(\vec{p}\,)\,
\left [~{T}^{(I=0)}_{\bar K^*N}(P_0,\vec{P})+3 {T}^{(I=1)}_{\bar K^*N}(P_0,\vec{P})\right ] .
 \label{eq:pid}
\end{eqnarray}
 The self-energy $\Pi_{\bar{K}^*}^{{\rm (b)}}(q_0,\vec{q}\,)$ has to be determined self-consistently since it is obtained from the in-medium amplitude 
${T}_{\bar K^*N}$ which contains the $\bar K^*N$ loop function
and this quantity itself is a function of
$\Pi_{\bar K^*}=\Pi_{\bar{K}^*}^{{\rm (a)}}
+\Pi_{\bar{K}^*}^{{\rm (b)}}$. 

\begin{figure}[t]
\begin{center}
\includegraphics[width=0.45\textwidth,height=6cm]{spectral_ksn.eps}
\hfill
\includegraphics[width=0.5\textwidth,height=5.5cm]{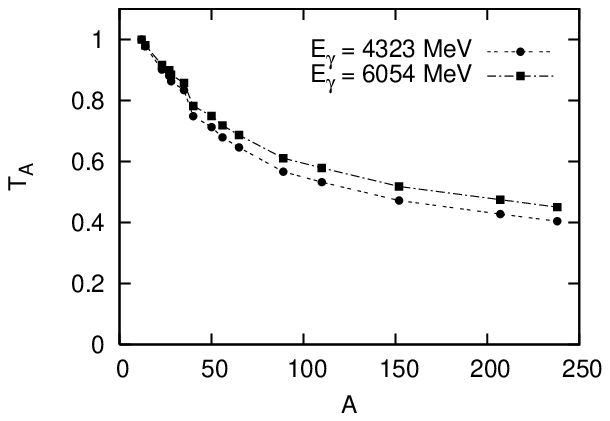}
\caption{ Left: $\bar K^*$ spectral function. Right: Transparency ratio for  $\gamma A \to K^+ K^{*-} A'$}
\label{fig:spec-trans}
\end{center}
\end{figure}

The $\bar K^*$ meson spectral function, which results from the imaginary part of the in-medium $\bar K^*$ propagator, is displayed in l.h.s. of Fig.~\ref{fig:spec-trans} as a function of the meson energy $q_0$, for zero momentum and different densities up to 1.5 $\rho_0$. The dashed line refers to the calculation in free space, where only the $\bar K \pi$ decay channel contributes, while the other three lines correspond to fully self-consistent calculations, which also incorporate the process $\bar K^* \rightarrow \bar K \pi$ in the medium. 

 The structures present above the quasiparticle peak correspond to $\Lambda(1783) N^{-1}$ and  $\Sigma(1830) N^{-1}$ excitations. Density effects result in a dilution and merging of those resonant-hole states, together with a general broadening of the spectral function  due to the increase of collisional and absorption processes. 
Although the real part of the optical potential is moderate, -50 MeV at $\rho_0$, the interferences with the resonant-hole modes push the $\bar{K}^*$  quasiparticle peak to lower energies. However, including the transitions to
pseudoscalar-meson states, such as $\bar K^*N \to \bar K N$, would make the peak
less prominent and difficult to disentangle from the other excitations.
In any case, what is clear from the present approach, is that the width
of the $\bar K^*$ increases substantially in the medium, becoming at normal nuclear
matter density five times bigger than in free
space.

\subsection{Transparency ratio for  $\gamma A \to K^+ K^{*-} A'$}

 In order to test experimentally the $\bar K^*$ self-energy, we can study the nuclear transparency ratio by comparing the cross sections of the photoproduction reaction $\gamma A \to K^+ K^{*-} A'$ in different nuclei, and tracing them to the in medium $K^{*-}$ width.

The normalized nuclear transparency ratio is defined as
\begin{equation}
T_{A} = \frac{\tilde{T}_{A}}{\tilde{T}_{^{12}C}} \hspace{1cm} ,{\rm with} \ \tilde{T}_{A} = \frac{\sigma_{\gamma A \to K^+ ~K^{*-}~ A'}}{A \,\sigma_{\gamma N \to K^+ ~K^{*-}~N}} \ .
\end{equation}
The quantity $\tilde{T}_A$ is the ratio of the nuclear $K^{*-}$-photoproduction cross section
divided by $A$ times the same quantity on a free nucleon. It describes the loss of flux of $K^{*-}$ mesons in the nucleus and is related to the absorptive part of the $K^{*-}$-nucleus optical potential and, thus, to the $K^{*-}$ width in the nuclear medium.  We evaluate the ratio between the nuclear cross sections in heavy nuclei and a light one ($^{12}$C), $T_A$, so that other nuclear effects not related to the absorption of the $K^{*-}$ cancel.

The results for different nuclei can be seen in the r.h.s of Fig. \ref{fig:spec-trans}, where the transparency ratio has been plotted for two different energies in the center of mass reference system $\sqrt{s}=3$ GeV and $3.5$ GeV, which are equivalent to energies of the photon in the lab frame of $4.3$ GeV and $6$ GeV respectively. We observe a very strong attenuation of the $\bar{K}^*$ survival probability due to the decay or absorption channels $\bar{K}^*\to \bar{K}\pi$ and $\bar{K}^*N\to \bar K^* N, \rho Y, \omega Y, \phi Y, \dots$ with increasing nuclear-mass number $A$. This is due to the larger path that the $\bar{K}^*$ has to follow before it leaves the nucleus, having then more chances to decay or get absorbed.

\section*{Acknowledgments}
L.T. acknowledges support from the RFF program of the University of Groningen. This work is partly supported by the EU contract No. MRTN-CT-2006-035482 (FLAVIAnet). We acknowledge the support of the European Community-Research Infrastructure Integrating Activity ``Study of Strongly Interacting Matter'' (HadronPhysics2, Grant Agreement n. 227431) under the 7th Framework Programme of EU.


\begin{thebibliography}{99}
  





\bibitem{Fuchs:2005zg}
C.~Fuchs,
Prog.\ Part.\ Nucl.\ Phys.  {\bf 56} (2006), 1.
\bibitem{fair} http://www.gsi.de/fair

\bibitem{Friedman:2007zz}
E.~Friedman and A.~Gal,
  Phys.\ Rept.  {\bf 452} (2007), 89. 
\bibitem{Koch}
 V.~Koch,
 Phys.\ Lett.\ B {\bf 337} (1994), 7. 
\bibitem{Lutz}
  M.~Lutz,
 Phys.\ Lett.\ B {\bf 426}  (1998), 12. 
\bibitem{Ramos:1999ku}
A.~Ramos and E.~Oset,
Nucl.\ Phys.\ A {\bf 671} (2000), 481.
\bibitem{Tolos01}
 L.~Tolos, A.~Ramos, A.~Polls and T.~T.~S.~Kuo,
 Nucl.\ Phys.\ A {\bf 690} (2001), 547.
\bibitem{Tolos02}
 L.~Tolos, A.~Ramos and A.~Polls, 
 Phys.\ Rev.\  C {\bf 65} (2002), 054907. 

\bibitem{Tolos:2006ny}
 L.~Tolos, A.~Ramos and E.~Oset, 
Phys.\ Rev.\  C {\bf 74} (2006), 015203. 

\bibitem{Lutz:2007bh}
 M.~F.~M.~Lutz, C.~L.~Korpa and M.~Moller, 
Nucl.\ Phys.\  A {\bf 808} (2008), 124. 

\bibitem{Tolos:2008di}
 L.~Tolos, D.~Cabrera and A.~Ramos, 
Phys.\ Rev.\  C {\bf 78} (2008), 045205. 

\bibitem{rapp}
  R.~Rapp and J.~Wambach,
  Adv.\ Nucl.\ Phys.\  {\bf 25} (2000), 1. 

\bibitem{hayano}
  R.~S.~Hayano and T.~Hatsuda,
  arXiv:0812.1702 [nucl-ex].

\bibitem{mosel}
  S.~Leupold, V.~Metag and U.~Mosel,
Int.\ J.\ Mod.\ Phys.\  E {\bf 19} (2010), 147.  


\bibitem{GarciaRecio:2005hy}
C.~Garcia-Recio, J.~Nieves, and L.~L.~Salcedo,
Phys.\ Rev. \ D {\bf 74} (2006), 034025.

\bibitem{Oset:2009vf}
  E.~Oset and A.~Ramos,
Eur.\  Phys. \ J. A {\bf 44} (2010), 445.

\bibitem{tolos10}
L.~Tolos, R.~Molina, E.~Oset and A.~Ramos, arXiv:1006.3454 [nucl-th].


\bibitem{ewsr} D.~Cabrera, A.~Polls, A.~Ramos and L.~Tolos,
  Phys.\ Rev.\  C {\bf 80} (2009), 045201.

\end{thebibliography}
\end{document}